\documentclass{jpp}
\usepackage{graphicx}
\usepackage{natbib}
\usepackage{epstopdf, epsfig}
\usepackage{hyperref}
\usepackage{amssymb}
\usepackage{amsmath}
\usepackage[]{enumerate}
\usepackage{bm}
\usepackage{xcolor}
\newcommand{\mathematica}{Mathematica\textsuperscript{\textregistered} }

\newcommand{\me}{\mathrm{e}}

\newcommand{\dif}{\mathrm{d}}
\newcommand{\al}{\alpha}

\DeclareBoldMathCommand{\bV}{V}
\DeclareBoldMathCommand{\bv}{v}
\DeclareBoldMathCommand{\bF}{F}
\DeclareBoldMathCommand{\bg}{g}
\DeclareBoldMathCommand{\bl}{\ell}
\DeclareBoldMathCommand{\bu}{u}
\DeclareBoldMathCommand{\br}{r}
\DeclareBoldMathCommand{\bx}{x}
\DeclareBoldMathCommand{\bz}{z}
\DeclareBoldMathCommand{\bg}{g}
\DeclareBoldMathCommand{\bb}{b}
\DeclareBoldMathCommand{\be}{e}
\DeclareBoldMathCommand{\bs}{s}
\DeclareBoldMathCommand{\bA}{A}
\DeclareBoldMathCommand{\bB}{B}
\DeclareBoldMathCommand{\bC}{C}
\DeclareBoldMathCommand{\bD}{D}
\DeclareBoldMathCommand{\bE}{E}
\DeclareBoldMathCommand{\bI}{I}
\DeclareBoldMathCommand{\bJ}{J}
\DeclareBoldMathCommand{\bM}{M}
\DeclareBoldMathCommand{\bL}{L}
\DeclareBoldMathCommand{\bN}{N}
\DeclareBoldMathCommand{\bP}{P}
\DeclareBoldMathCommand{\bR}{R}
\DeclareBoldMathCommand{\bS}{S}
\DeclareBoldMathCommand{\bu}{u}
\DeclareBoldMathCommand{\bW}{W}
\DeclareBoldMathCommand{\bk}{k}
\DeclareBoldMathCommand{\ba}{a}
\DeclareBoldMathCommand{\bn}{n}
\DeclareBoldMathCommand{\bp}{p}
\DeclareBoldMathCommand{\bq}{q}
\DeclareBoldMathCommand{\br}{r}

\shorttitle{Symmetries of the ZS model}
\shortauthor{R. L. White, R. D. Hazeltine and N. F. Loureiro}

\title{Symmetries of a reduced fluid-gyrokinetic system}

\author{R. L. White\aff{1}
  \corresp{\email{rlw@mit.edu}},
  R. D. Hazeltine\aff{2}
 \and N. F. Loureiro\aff{1}}

\affiliation{\aff{1}Plasma Science and Fusion Center, Massachusetts Institute of Technology,
Cambridge, MA 02139, USA
\aff{2}Institute for Fusion Studies, University of
Texas at Austin, Austin TX 78712}

\begin{document}

\maketitle

\begin{abstract}
Symmetries of a fluid-gyrokinetic model are investigated using Lie group techniques.  Specifically the nonlinear system constructed by Zocco and Schekochihin \cite[]{zoccoetal}, which combines nonlinear fluid equations with a drift-kinetic description of parallel electron dynamics, is studied.  Significantly, this model is fully gyrokinetic, allowing for arbitrary $k_{\perp}\rho_{i} $, where $k_{\perp} $ is the perpendicular wave vector of the fluctuations and $\rho_{i}$ the ion gyroradius.  The model includes integral operators corresponding to gyroaveraging as well as the moment equations relating fluid variables to the kinetic distribution function.  A large variety of exact symmetries is uncovered, some of which have unexpected form.  Using these results, new nonlinear solutions are constructed, including a helical generalization of the Chapman-Kendall solution for a collapsing current sheet.
\end{abstract}

\section{Introduction}\label{intro}
Symmetry transformations -- changes in the de\-pen\-dent and in\-de\-pen\-dent vari\-ables of a phy\-sical model that leave the model equations unchanged -- are revealing and useful throughout theoretical physics.  The most general scheme for uncovering point symmetries of a system of equations is Lie group analysis \cite[see, e.g.,][]{olver, cantwell}.  This scheme has been used extensively in plasma physics, including studies of the Vlasov-Maxwell model for an unmagnetized plasma \cite[see][]{Roberts85, kovalev} and the Grad-Shafranov equation \cite[]{SymGS}. A special case of Lie symmetry, scaling symmetry, was fruitfully employed by \cite{ConnnorTaylor1977}. In this work we apply the Lie procedure to a particular nonlinear gyrokinetic fluid model used in magnetized plasma turbulence and magnetic reconnection studies. 

The symmetries of any physical model have intrinsic interest, especially because one often uncovers unexpected symmetries -- beyond the usual rotations, translations and so on which are obvious from physical considerations.  Knowledge of the symmetries can simplify numerical calculations, while providing useful tests on their accuracy.  When a variational principle is available, the symmetries can be used to identify dynamical constants.  They can also be used to generate new solutions from old ones -- in particular, physically interesting solutions can be constructed by applying the group operator to a trivial, less interesting solution.  Finally, in many cases symmetries can be used to reduce the order of a differential equation system, in some cases leading to exact solutions. 

\subsection{Fluid-gyrokinetic model}
A magnetized plasma is one in which the ion gyro-radius, $\rho_{i}$, is small compared to all equilibrium gradient scale lengths. But scale lengths of \emph{perturbed} quantities in a magnetized plasma, measured by the perpendicular wave length $k_{\perp}^{-1}$, can break this ordering:  $k_{\perp}\rho_{i} \sim 1$.  Theories allowing for such finite-Larmor-radius (FLR) effects increasingly dominate plasma physics research, entering both kinetic and fluid models of plasma dynamics.

There are two ways in which conventional fluid equations fall short in their description of magnetized plasma dynamics. First, they represent FLR effects crudely, retaining at most terms of second order in $k_{\perp} \rho_{i}$.  Second, they entirely omit Landau resonances, which, in the magnetized context, enter through wave-particle interactions parallel to the field -- effects conventionally treated by the drift-kinetic equation.  \emph{Gyrokinetics}~\cite[]{RuthFr,TaylorHastie, CattoI,CattoII,FriemanChen,Dubin,LeeI,LeeII,HahmLee,Brizard92} addresses both shortcomings, providing in particular a full FLR treatment of the perturbed fields, with however the expense and complexity of computation (analytical and numerical) in five dimensions of phase space.  \emph{Gyro-fluid} models reduce this overhead by restricting the FLR physics to coordinate space \cite[see, for example,][]{HammettPerkins, Hammett92, DorlandHammett93, HammettEtal93, Beer96, Snyder2001, flw09, bian}. However, the validity of the approximations made in their derivation can be hard to ascertain, especially in nonlinear contexts~\citep{Dimits}. 

An alternative and conceptually straightforward approach combines a fluid treatment of the perpendicular physics with a drift-kinetic description, including resonances and collisions, of the parallel dynamics \cite[]{RamosI, RamosII}.   Such a hybrid approach was proposed and applied as early as 1958 \cite[]{kruskaloberman, rosenbluth-rostoker}. Called ``kinetic MHD,''  the early approach neglected most FLR effects, combining MHD with the drift-kinetic equation.  However in other respects it resembles the gyrokinetic fluid hybrid considered here.

We study a particular representative of the fluid-kinetic approach: the reduced gyrokinetic model derived in \cite{zoccoetal}, referred to below as ZS.  The model uses five fields -- five functions of five independent variables (including time).  To make this work self-contained, and establish notation, we start by reviewing the physical assumptions built into the ZS model in Section \ref{zsn}, and then express the model equations in normalized variables in Section \ref{normeqns}.  The remainder of the paper uses exclusively normalized variables, so the reader who is already familiar with the model can skip Section \ref{zsn}. The symmetries obtained from our analysis are shown in Section \ref{symmetries_section}.  Section \ref{app_section} presents new exact solutions of the reduced MHD (RMHD) equations--a limit of the ZS model--obtained using symmetry transformations.  In Section \ref{analysis} we display the Lie group generator and present the procedure used to derive the symmetries for the (integro-differential) ZS model.  We do not attempt any full exegesis of the Lie procedure; readers unfamiliar with it may consult such texts as \cite{olver} or \cite{cantwell}.  Our conclusions are summarized and discussed in Section \ref{con}.

\section{Model equations} \label{mod}

\subsection{Introduction}\label{zsn}
The detailed derivation of the ZS model from the gyrokinetic equations is given in \cite{zoccoetal}.  Here, we briefly survey the physical assumptions, summarize the resulting equations, and indicate the physical meaning of each of the fields.  

The plasma, composed of electrons with charge $-e$ and ions with charge $Ze$, is assumed to have a uniform background magnetic field $B_0\hat{\mathbf{z}}$, and the equilibrium electrons and ions are Maxwellian:
\begin{equation}
F_{0s} = \frac{n_{0s}}{(2\pi)^{3/2}v_{Ts}^3}\text{exp}\left(-\frac{v^2}{2v_{Ts}^2}\right),
\end{equation}
with $v_{Ts}=(T_{0s}/m_s)^{1/2}$.  Here we deviate from the convention of \cite{zoccoetal}, where the Maxwellian is characterized by its \emph{most probable} speed $v_{\text{th},s}=\sqrt{2}v_{Ts}$.  This translates to a slightly different definition of the Larmor radius, which for us is defined $\rho_s =v_{Ts}/|\Omega_s|$, with $\Omega_s = q_s B_0/(m_s c)$.  This modification eliminates many factors of $\sqrt{2}$ in the final equations.  

In accordance with the standard $\delta f$ gyrokinetic ansatz, each field is split into its background value plus a small perturbation, with $\delta f_s/F_{0s}\sim \delta \mathbf{B}/\mathbf{B}_0 \sim k_{\parallel}/k_{\perp} \sim \omega/\Omega_s \sim \epsilon \ll 1$, and, additionally, $\beta_s \sim Z m_e/m_i$, with the mass ratio being treated as a second formal small parameter.

\subsubsection{Electrostatic Ions}
After ordering out electromagnetic effects and parallel streaming in the ion gyrokinetic equation, the ion distribution function is approximated by
\begin{equation} \label{ion_delta_f}
\delta f_i = \frac{ZeF_{0i}}{\tau T_{0e}}\left(\langle \varphi \rangle_{\mathbf{R}_i}-\varphi \right),
\end{equation} 
where $\mathbf{R}_i(\mathbf{r},\mathbf{v})=\mathbf{r}-\Omega_{i}^{-1}\mathbf{v}\times\hat{\mathbf{z}}$, $\Omega_{i}=Ze\mathbf{B}_0/(m_ic)$, $\tau = T_{0i}/T_{0e},$ and $\langle \cdots \rangle_{\mathbf{R}_{i}}$ denotes the gyroaverage at fixed $\mathbf{R}_i$.

It follows that the ion density perturbation $\delta n_{i}$ and mean parallel flow $u_{\parallel,i}$ are given by
\begin{align}
\frac{\delta n_i}{n_{0i}}=&-\frac{Z}{\tau}(1-\hat{\Gamma}_0)\frac{e\varphi}{T_{0e}},\label{ion_delta_n}\\
u_{\parallel,i}=&0, \label{static_ions}
\end{align}
where $\hat{\Gamma}_0$ is an ion gyroaveraging operator:
\begin{align}
\hat{\Gamma}_0[\cdots] \equiv \frac{1}{n_{0i}}\int\text{d}^3\mathbf{v}\langle\langle \cdots \rangle_{\mathbf{R}_i} \rangle_{\mathbf{r}}F_{0i}(\mathbf{v}).
\end{align}
In Fourier space, $\Gamma_{0}$ has the closed-form expression
\begin{align} \label{GammaFourier}
\Gamma_{0} = I_0(\alpha_i)\text{e}^{-\alpha_i},
\end{align}
where $I_0$ is the zeroth order modified Bessel function and $\alpha_i =k_{\perp}^2\rho_i^2$. 

\subsubsection{Quasineutrality and Amp\`ere's Law}
Since $u_{\parallel,i}=0$, we have $J_{\parallel} =-en_{0e}u_{\parallel e}$, and thus the parallel component of Amp\`ere's law becomes
\begin{equation} \label{amperes_law}
u_{\parallel,e}=\frac{e}{m_e c}d_e^2\nabla_{\perp}^2A_{\parallel}.
\end{equation}
 
According to Eq. (\ref{ion_delta_n}), quasineutrality is expressed by
\begin{equation}
\frac{\delta n_e}{n_{0e}}=-\frac{Z}{\tau}(1-\hat{\Gamma}_0)\frac{e\varphi}{T_{0e}}\label{electron_delta_n}.
\end{equation}

\subsubsection{Drift-kinetic electrons}
%At low $\beta$, both electron FLR effects and the $\nabla B$ drift can be neglected, leading to a drift kinetic equation which, in the absence of collisions, has no explicit $v_{\perp}$ dependence.  After making the first two moments from the electron distribution function explicit,
The electrons are described by a distribution function $g_e$ from which the density and parallel flow terms have been extracted:
\begin{equation}
g_e = \delta f_e-\left(\frac{\delta n_e}{n_{0e}}-v_{\parallel}\frac{u_{\parallel,e}}{v_{Te}^2}\right) F_{0e}, \label{defg1}
\end{equation}
The electron dynamics is described by fluid equations for the explicit moments, together with a simplified drift kinetic equation:
\begin{align}
&\frac{\mathrm{d}}{\mathrm{d}t}\frac{\delta n_e}{n_{0e}} =-\hat{\mathbf{b}}\cdot\nabla u_{\parallel e},\label{electron_zeroth_moment}\\
&\frac{\mathrm{d}}{\mathrm{d}t}\left(A_{\parallel}+\frac{c m_e}{e}u_{\parallel e} \right)=-c\frac{\partial \varphi}{\partial z}+\frac{cT_{0e}}{e}\hat{\mathbf{b}}\cdot\nabla \left(\frac{\delta n_{e}}{n_{0e}}+\frac{\delta T_{\parallel,e}}{T_{0e}} \right),\label{electron_first_moment}
\end{align}
where
\begin{align}
\frac{\delta T_{\parallel,e}}{T_{0e}}=&\frac{1}{n_{0e}}\int \text{d}^3 \mathbf{v} \frac{v_{\parallel}^2}{v_{Te}^2}g_e,
\end{align}
is the electron temperature perturbation.  We have also introduced the convective time derivative
\begin{align}
\frac{\mathrm{d}f}{\mathrm{d}t}\equiv& \frac{\partial f}{\partial t}+\frac{c}{B_0}\left\{\varphi,f\right \},
\end{align}
with the Poisson bracket defined by
\begin{align} \label{bracket_def}
\{f,g\} \equiv \frac{\partial f}{\partial x} \frac{\partial g}{\partial y}- \frac{\partial f}{\partial y}\frac{\partial g}{\partial x},
\end{align}
and the parallel gradient operator
\begin{align}
\hat{\mathbf{b}}\cdot\nabla f\equiv& \frac{\partial f}{\partial z}-\frac{1}{B_0}\left\{A_{\parallel},f\right \}.
\end{align}
The remaining distribution $g_{e}$ is determined by a simplified drift-kinetic equation:
\begin{equation}
\begin{split}
\frac{\mathrm{d}g_e}{\mathrm{d}t} &+ v_{\parallel}\hat{\mathbf{b}}\cdot\nabla\left(g_e-\frac{\delta T_{\parallel,e}}{T_{0e}}F_{0e} \right)-C[g_e]\\
&=\left(1-\frac{v_{\parallel}^2}{v_{Te}^2} \right)F_{0e}\hat{\mathbf{b}}\cdot\nabla u_{\parallel,e},\label{electron_g_eqn}
\end{split}
\end{equation}
in which electron FLR terms, as well as curvature drifts, are ordered out by the strong guide field.
 Finally, 
\begin{equation} \label{def_coll_int}
\begin{split}
C[g_e]=\nu_{ei}\Bigg[&v_{Te}^2\frac{\partial}{\partial v_{\parallel}}\left(\frac{\partial}{\partial v_{\parallel}}+\frac{v_{\parallel}}{v_{Te}^2} \right)g_e-\left(1-\frac{v_{\parallel}^2}{v_{Te}^2} \right)\frac{\delta T_{\parallel e}}{T_{0e}}F_{0e} \Bigg]
\end{split}
\end{equation}
is a model collision operator that conserves particles, parallel momentum and parallel kinetic energy \cite[]{zoccoetal}.  This model operator -- a generalization of the so-called Lenard-Bernstein operator introduced by \cite{ray} -- also satisfies an $H$ theorem.

Note that (\ref{defg1}) requires the integral constraints
\begin{equation}\label{int_constraints}
\int \text{d}^3\mathbf{v}\left(\begin{array}{c}1\\ v_{\parallel}\end{array} \right)g_e=0.
\end{equation}

\subsubsection{Summary}
Given a background characterized by $B_0$ and $v_{Ts}$, Eqs. (\ref{amperes_law}), (\ref{electron_delta_n}), (\ref{electron_zeroth_moment}), (\ref{electron_first_moment}), and (\ref{electron_g_eqn}), are a closed system of equations governing small nonlinear perturbations of the fields $\varphi$, $A_{\parallel}$, $u_{\parallel e}$, $\delta n_e/n_{0e}$, and $g_e$.  In the final formulation of the model presented in Eqs. (62)--(64) of~\cite{zoccoetal}, $u_{\parallel e}$ has been eliminated using (\ref{amperes_law}).  We will do the same in the remainder of the paper.

\subsection{Normalization} \label{normeqns}
For the purposes of obtaining symmetries, it is convenient to reduce the number of constants in the ZS model by normalizing all quantities.  It turns out that the fields can be normalized in such a way that there are only two dimensionless constants: $Z/\tau$ and $\alpha \equiv \rho_{i}^{2}/d_{e}^{2}$, and these only appear in the integral closure relation relating the electrostatic potential to the density perturbation.

The dependent variables are normalized via
\begin{equation}
\begin{array}{ccc}
\delta n=\frac{\delta n_e/n_{0e}}{\left\langle \delta n_e/n_{0e} \right\rangle},& \psi=\frac{A_{\parallel}}{\left\langle A_{\parallel} \right\rangle},& \phi = \frac{\varphi}{\langle \varphi \rangle},\\
\delta T = \frac{\delta T_{\parallel e}/T_{0e}}{\left\langle  \delta T_{\parallel e}/T_{0e}\right\rangle},& g = \frac{g_e}{F_{0e}\langle g_e\rangle},
\end{array}
\end{equation}
with
\begin{align}
\left\langle A_{\parallel} \right\rangle =& B_0 \nu_{ei}d_e^2/v_{Te},\\
\langle \varphi \rangle =& B_0 \nu_{ei}d_e^2/c,\\
\left\langle \delta n_e/n_{0e} \right\rangle =& \left\langle  \delta T_{\parallel e}/T_{0e}\right\rangle=\langle g_e\rangle=\langle\delta\rangle,
\end{align}
where
\begin{equation}
\langle \delta \rangle \equiv \frac{e B_0 \nu_{ei}d_e^2}{c T_{0e}} = \frac{\nu_{ei}}{\Omega_e}\sqrt{\beta_e}.
\end{equation}

The independent variables are similarly normalized, with the following normalization scales:
\begin{equation}
\begin{array}{ccccc}
\langle v_{\parallel}\rangle = v_{Te},& \langle x_{\perp} \rangle = d_e,& \langle z\rangle = v_{Te}/\nu_{ei},& \langle t\rangle = 1/\nu_{ei}.
\end{array}
\end{equation}

Defining the normalized convective time derivative, parallel gradient, and perpendicular Laplacian
\begin{equation}
\begin{array}{ccc}
\text{d}_t f \equiv \p_{t}f + \{\phi, f\},& \nabla_{\parallel}f \equiv \p_{z}f - \{\psi,f\},& \Delta \equiv \partial_x^2 + \partial_y^2,
\end{array}
\end{equation}
and normalized gyrokinetic and collision operators
\begin{align}
\hat{\mathcal{G}} \equiv& -\frac{Z}{\tau}\frac{e\langle \varphi\rangle}{\langle \delta \rangle}\mathcal{G},\\
\hat{C} \equiv&g_{vv}-vg_v-(1-v^2) \delta T,
\end{align}
the normalized reduced fluid-kinetic model takes the form
\begin{align}
\text{d}_t\delta n =& -\nabla_{\parallel}\Delta \psi,\label{eqn1}\\
\text{d}_t\psi + \phi_z  =& {\lambda}\left[\Delta (\psi+d_t\psi) +\nabla_{\parallel}(\delta n + \delta T)\right],\label{eqn2}\\
\text{d}_tg + v\nabla_{\parallel}(g - \delta T) =&  \hat{C} + (1 - v^2)\nabla_{\parallel}\Delta \psi ,\label{eqn3}\\
\delta T =& \frac{1}{\sqrt{2\pi}}\int\mathrm{d}v' v'^2 \mathrm{e}^{-{v'}^2/2}g(v'),\label{closure_eqn1}
\end{align}
Here, the brackets are the same as (\ref{bracket_def}) except the perpendicular coordinates are now normalized, and $\lambda(=1)$ is a tag for the terms that are dropped in the ideal reduced magnetohydrodynamic (RMHD) limit.  These differential equations are to be solved subject to the integral constraints
\begin{align} \label{norm_int_constraints}
\left(\begin{array}{c}0\\0\end{array}\right) =& \frac{1}{\sqrt{2\pi}}\int\mathrm{d}v' \left(\begin{array}{c}0\\v'\end{array}\right) \mathrm{e}^{-{v'}^2/2}g(v'),
\end{align}
together with
\begin{equation}
 \delta n = \hat{\mathcal{G}}\phi\label{closure_eqn2}.
\end{equation}
We introduce a normalized Alfv\`en velocity,
\begin{align}
v_{A}=&\frac{1}{v_{Te}}\frac{B_0}{\sqrt{4\pi n_{0i}m_i}}\\
(=&\sqrt{\tau \al/Z}),
\end{align}
and the normalized kernel,
\begin{align}
\label{defk}
\hat{K}(x) =& \frac{-Z/\tau}{2 \pi} \int \dif k_{\perp} \, k_{\perp}J_{0}(k_{\perp} x)\left[1 - I_{0}(\alpha k_{\perp}^{2})\me^{-\alpha k_{\perp}^{2}}\right]
\end{align}
obtained from (\ref{GammaFourier}).  Then the operator $\hat{\mathcal{G}}$ becomes
\begin{align}
\hat{\mathcal{G}}u = &\int\mathrm{d}^2x_{\perp}'\hat{K}(|\mathbf{x}_{\perp}-\mathbf{x}_{\perp}'|)u(\mathbf{x}_{\perp}'),\label{gyro_integral}\\
		  = &v_A^{-2} \Delta + \lambda\mathcal{O}(\alpha^2 \Delta^2).\label{FLRgyro}
\end{align}

Finally, to determine the symmetries of these equations, we must explicitly include the trivial relations 
\begin{align}
\partial_v \delta n = \partial_v \phi=\partial_v \psi=\partial_v \delta T=0. 
\end{align}
\subsection{RMHD Limit}
If we set  $\lambda = 0$ in (\ref{eqn2}) and (\ref{FLRgyro}), the equations (\ref{eqn1})--(\ref{eqn2}) become an autonomous subsystem for $\phi$ and $\psi$:
\begin{align}
\text{d}_t\Delta\phi =& -v_A^2\nabla_{\parallel}\Delta \psi,\label{RMHDeqn1}\\
\text{d}_t\psi + \phi_z  =& 0,\label{RMHDeqn2}
\end{align}
while $g$ becomes a decoupled scalar field, constrained to satisfy the driven integro-differential equation
\begin{equation}
\text{d}_tg + v\nabla_{\parallel}(g - \delta T[g]) =  g_{vv}-vg_v + (1 - v^2)(\nabla_{\parallel}\Delta \psi-\delta T[g]).
\end{equation}

Equations (\ref{RMHDeqn1}) and (\ref{RMHDeqn2}) define ideal RMHD \cite[]{kadomtsev,strauss}.  There is no coupling to the kinetic equation.

\section{Symmetries} \label{symmetries_section}
Here, we present the symmetries of the system (\ref{eqn1})--(\ref{closure_eqn2}) in the form of transformations of known solutions, rather than in terms of the infinitesimal generators of the symmetries.  The latter  are obtained directly from the invariance criterion in section \ref{analysis}.

\subsection{Gauge Transformation}
Given a solution $(\phi, \psi, \delta n, \delta T, g)(v,\mathbf{x}_{\perp},z,t)$, and an arbitrary function $H(z,t)$, one can generate a new solution $(\tilde{\phi}, \tilde{\psi}, \tilde{\delta n}, \tilde{\delta T}, \tilde{g})(v,\mathbf{x}_{\perp},z,t)$ via
\begin{equation} \label{norm_gauge_trans}
\left(\begin{array}{c}\tilde{\phi}\\ \tilde{\psi}\\ \tilde{\delta n}\\ \tilde{\delta T}\\ \tilde{g}\end{array}\right)(v,\mathbf{x}_{\perp},z,t) = \left(\begin{array}{c}\phi \\ \psi\\ \delta n \\ \delta T\\ g\end{array}\right)(v,\mathbf{x}_{\perp},z,t) + \left(\begin{array}{c}-\partial_t H\\ \partial_z H\\0\\0\\0  \end{array} \right). 
\end{equation} 

It is not hard to see that this symmetry is expressing gauge invariance.  After undoing the normalizations, (\ref{norm_gauge_trans}) becomes
\begin{equation}
\left(\begin{array}{c}\tilde{\varphi}\\ \tilde{\mathbf{A}}\end{array}\right) = \left(\begin{array}{c}\varphi \\ \mathbf{A}\end{array}\right) + \left(\begin{array}{c}-\frac{1}{c}\frac{\partial \Lambda}{\partial t}\\ \nabla \Lambda  \end{array} \right), 
\end{equation} 
where $\Lambda = c\langle \varphi \rangle\langle t\rangle H$, and $\mathbf{A}=\hat{\mathbf{z}}A_{\parallel}+\mathcal{O}(\sqrt{\beta_s})$.  Note that if $H$ had $\mathbf{x}_{\perp}$ dependence, then this gauge transformation would change $\mathbf{A}_{\perp}$ as well.  However, in the low-$\beta$ limit, $\mathbf{A}_{\perp}$ is ordered out of the model, so the gauge must be independent of $\mathbf{x}_{\perp}$.

Of course this symmetry also holds in the RMHD model -- explaining the absence of $\lambda$ in the transformation (\ref{norm_gauge_trans}).

The appearance of gauge symmetry in the ZS model is not surprising, but also not without significance.  In view of the many approximations involved in the construction of ZS and other reduced models, its emergence here gives confidence in the model's treatment of the electromagnetic field.

\subsection{Perpendicular Translations}
Let $\boldsymbol{\xi}(z,t)$ be an arbitrary displacement in the $x$-$y$ plane.  Then $\boldsymbol{\xi}$ produces the symmetry transformation 
\begin{equation}\label{perp_translations}
\left(\begin{array}{c}\tilde{\phi}\\ \tilde{\psi}\\ \tilde{\delta n}\\ \tilde{\delta T}\\ \tilde{g}\end{array}\right)(v,x,y,z,t) = \left(\begin{array}{c}\phi \\ \psi\\ \delta n \\ \delta T\\ g\end{array}\right)(v,\mathbf{x}_{\perp}+\boldsymbol{\xi}(z,t),z,t) + \left(\begin{array}{c}-\partial_t \left(\hat{\mathbf{z}}\cdot \boldsymbol{\xi}\times \mathbf{x}_{\perp}\right)\\ \partial_z (\hat{\mathbf{z}}\cdot \boldsymbol{\xi}\times \mathbf{x}_{\perp})\\0\\0\\0  \end{array} \right). 
\end{equation} 
In the case where $\boldsymbol{\xi}$ is a constant, we recover the obvious result that the model is translation invariant in the $\mathbf{x}_{\perp}$ plane.  In the more general case, the transformations of $\phi$ and $\psi$ follow the same pattern as the gauge symmetry, but the overall transformation of these fields is not a gauge transformation: note the additional $(z,t)$-dependent translation of the initial fields, as well as the fact that the gradient of $\hat{\mathbf{z}}\cdot \boldsymbol{\xi}\times \mathbf{x}_{\perp}$ has nonzero $\hat{\mathbf{x}}_{\perp}$ components.

\subsection{Alfv\`enic Rotations}\label{alf_rotations}
Let $\Theta(z,t)$ be a solution to the one dimensional (Alfv\`en) wave equation 
\begin{equation}\label{Theta_eqn}
v_A^{-2}\Theta_{tt}=\Theta_{zz}.
\end{equation}
This function will determine the $z$-$t$ dependent rotation of the original solution about the $z$ axis.  After transforming to polar coordinates in the $x$-$y$ plane, $\mathbf{x}_{\perp} = r\hat{\mathbf{r}}(\theta)$, the symmetry transformation takes the form
\begin{equation}
\begin{split} \label{alfven_rotations}
\left(\begin{array}{c}\tilde{\phi}\\ \tilde{\psi}\\ \tilde{\delta n}\\ \tilde{\delta T}\\ \tilde{g}\end{array}\right)(v, r,\theta,z,t) =& \left(\begin{array}{c}\phi \\ \psi\\ \delta n \\ \delta T\\ g\end{array}\right)(v,r,\theta+\Theta,z,t) + \left(\begin{array}{c}-\partial_t (r^2\Theta/2)\\ \partial_z (r^2\Theta/2)\\-2\partial_t\Theta/v_A^2\\0\\ G   \end{array} \right)+\lambda\left(\begin{array}{c} \mathcal{F}\\ 0\\0\\\mathcal{T}\\ 0  \end{array} \right) .
\end{split}
\end{equation} 
Here the function $G(v,z,t)$ appears as a displacement for the distribution function $g$:
\begin{equation}
g \rightarrow g + G.
\end{equation}
A detailed discussion of $G$ appears in the  following subsection.  We have also introduced
\begin{align}\label{phi_ext}
\mathcal{F} =   2(\Theta+\partial_t\Theta) -2\partial_t\Theta/v_A^{2}+\mathcal{T},
\end{align}
and
\begin{equation} \label{G2inT}
\mathcal{T} =  \frac{1}{\sqrt{2\pi}}\int\mathrm{d}v' v'^2 \mathrm{e}^{-{v'}^2/2}G(v').
\end{equation}

The first term on the right side of (\ref{phi_ext}) is due to the combination of resistivity and electron inertia; the second term arises from the density contribution to the perturbed electron pressure; and the last term is due to the electron temperature perturbation. 

Note that our symmetries apply to limiting case of RMHD, where the functions $G$ and $\mathcal{T}$ can be ignored.  

%It is worth noting here that the transformation involves $\delta n = -v_A^{-2}\Delta \phi$ in addition to the dependent variables $\phi$ and $\psi$ directly.  Transformations that explicitly involve derivatives of the dependent variables are technically outside the scope of classical point groups, and fall in the more general category of Lie-Backlund transformations.  
%
%Note that if we constrain the transformation of the derivative to vanish, as would be the case if we had considered the symmetries of RMHD directly, then $\Theta$ would have to be a constant, which is much less interesting than the more general case obtained here.  

\subsection{The function $G$}
\subsubsection{Linear drift-kinetic equation}
The function $\Theta$ determines $G$ implicitly, through the kinetic equation
\begin{equation}
G_{vv} - vG_v - v G_z -  G_t = -\text{He}_2(v)(\mathcal{T}-2\Theta_{tt}/v_A^2) - \text{He}_1(v) \mathcal{T}_{z},\label{G_diff_eqn}
\end{equation}
with the constraints
\begin{equation}
\left(\begin{array}{c}0\\0\end{array}\right) = \frac{1}{\sqrt{2\pi}}\int\mathrm{d}v' \left(\begin{array}{c}0\\v'\end{array}\right) \mathrm{e}^{-{v'}^2/2}G(v').\label{int_const_G}
\end{equation}
In (\ref{G_diff_eqn}), the $\text{He}_n$ are the ``probabilist's" Hermite polynomials.
  
Aside from the coefficients on its right-hand side, (\ref{G_diff_eqn}) is identical to the \emph{linearized} version of the drift-kinetic equation (\ref{electron_g_eqn}), which has been previously studied in detail~\cite[see, for example, ][]{zoccoetal, hatch2014,scheko2016,rr}.  In the symmetry context, the linearity of (\ref{G_diff_eqn}) does not result from approximation; here the linearity follows from the general structure of Lie groups.  In particular, the infinitesimal generators of any Lie group form a vector space, so the determining equations for symmetry transformations are always linear.  Similarly the absence of the electrostatic potential in (\ref{G_diff_eqn}) is not an approximation; it reflects exact Lie symmetries, such as (\ref{dt1}), (\ref{dt2}) and (\ref{dt8}).   

\subsubsection{Closed form solution for $G(v,z,t)$} \label{sec:label}
Using special choices for such functions as 
\begin{equation}
\Theta(z,t) = \Theta_{+}(z + v_{A}t) + \Theta_{-}(z - v_{A}t),
\end{equation}
it is not hard to find a closed-form solution for $G$.  Here we are content to display a single example: the choice
\begin{equation}
\Theta = -\frac{T_{0}}{48}\left[(z + v_{A}t)^{3} + (z - v_{A}t)^{3}\right],
\end{equation}
where $T_{0}$ is a constant, together with
\begin{equation}
\mathcal{T}  = \frac{T_{0}}{4 v_{A}}\left[(z + v_{A}t)^{2} - (z - v_{A}t)^{2}\right] = T_{0}zt
\end{equation}
allows an exact solution with
\begin{equation}
\label{gex}
G(v,z,t) = \frac{T_{0}}{2}\left[tz\text{He}_2(v) - \frac{1}{3}\left(t - \frac{1}{3}+ c\,\me^{-3t}\right)\text{He}_3(v)\right],
\end{equation}
where $c$ is an arbitrary constant.  It is easily verified that this function satisfies the differential equation as well as the integral constraints.  In addition to furnishing an explicit symmetry, this relatively simple function is in fact an exact solution to the full nonlinear integrodifferential model, and thus can be used for benchmarking codes.  

\subsubsection{Fourier-Hermite expansion of $G$}
A conventional approach to the drift-kinetic equation \citep{ws2004, zoccoetal,hatch2014,kanekar_schekochihin_dorland_loureiro_2015} expands the distribution function, in this case $G(v,z,t)$, as a series in Hermite polynomials.  Here it is convenient to use ``probabilists" Hermite polynomials, and to Fourier analyze the $z$- and $t$-dependent Hermite coefficient, thus expressing $G$ as
\begin{equation} \label{FH_expansion}
G(v,z,t) = \sum_{0}^{\infty}\frac{\text{He}_n(v)}{\sqrt{n!}}\int\frac{\text{d}\omega\text{d}k}{(2\pi)^2}G_n(\omega,k)\text{e}^{\text{i}(k z-\omega t)}.
\end{equation}
We note that this expansion restricts our consideration to solutions which have a Fourier transform; it would exclude, for example, (\ref{gex}). 

The constraints (\ref{int_const_G}) become
\begin{equation}
G_0 = G_1 = 0,
\end{equation}
while (\ref{G2inT}) gives
\begin{equation}
G_2 = \frac{1}{\sqrt{2}}\mathcal{T}.
\end{equation}
The remaining $G_n$ are determined by the recursion relation  
\begin{equation} \label{Grecursion}
k\left(\sqrt{n + 1} G_{n + 1} + \sqrt{n} G_{n - 1}\right) -\text{i} n G_n -\omega G_n = 0, \,\,\,n>2.
\end{equation}

Although (\ref{Grecursion}) is a simple, linear recursion relation, solving it requires some care:  there is a spurious divergent ``solution''~\citep{rr}  that must be avoided by appropriate determination of initial data -- in this case the ratio $\Delta \equiv G_{3}/G_{2}$.  In numerical applications, one is only interested in calculating a finite subset $\{G_n \}_{n\le N}$ because $g$ is represented by a finite sum of Hermite polynomials.  In this case, $\Delta(N)$ can be determined by the same closure scheme adopted by the numerical method to solve the full nonlinear model.  For example, if one simply truncates by setting $G_{N+1}=0$, then (\ref{Grecursion}) can be iterated backward to determine $\{G_N/G_{N-1}, G_{N-1}/G_{N-2},\, \dots, \Delta(N)\}$.  See \cite{ Zocco2015,viriato} for alternate closure schemes. The choice of $\Delta$, as well as other approaches to solving the recursion relation, will be considered in a future publication.

Finally, $\mathcal{T}$ is expressed in terms of the $\Theta$ and $\Delta$ via
\begin{equation}\label{formula_deltaT}
\mathcal{T}(z,t) = \int \left(\frac{2\text{i} k }{\frac{\omega/k}{2}-\sqrt{3}\Delta(\omega,k)}\right)\Theta(\omega,k) \text{e}^{-\text{i}k z +\text{i}\omega t }\frac{\text{d}\omega\text{d}k}{(2\pi)^2}.
\end{equation}
This expression, like (\ref{Grecursion}), is obtained by direct Fourier-Hermite transformation of (\ref{FH_expansion}).
 
\section{Sample Applications} \label{app_section}
The value of knowing the symmetries of a some mathematical description is appreciated in nearly all areas of physics.  In addition to their relation to conservation laws (discussed below), symmetries can be used to test numerical solution schemes, to motivate approximation hypotheses, and to generate novel exact solutions. In an important sense, the symmetries of a system carry information about its deep structure. The following discussion, touching upon two samples of potential application, is merely intended to be suggestive.
  
\subsection{Transforming the trivial solution} \label{triv_trans}
The most straightforward way to use symmetry transformations is to generate new solutions from known solutions.  One obvious exact solution is the trivial solution, with all of the fields identically zero.  In this case, by specifying the functions $H$, $\boldsymbol{\xi}$, and $\Theta$, one can generate nontrivial exact solutions by transforming the trivial solution using the symmetries presented in Section \ref{symmetries_section}.  In fact, one can directly verify that all of the nonlinear terms in the set of solutions obtained this way are exactly zero.  In other words, by transforming the trivial solution, one obtains solutions to the linearized version of the model which happen to be exact solutions to the full nonlinear system.   
\subsection{Transformed Chapman-Kendall solution}\label{trans_ck}
As a second illustration of the use of the transformations presented in section \ref{symmetries_section}, we consider the exact solution
\begin{equation}\label{ck_soln}
\begin{array}{cc}\phi = \Gamma x y,& \psi = \frac{x^2}{a_0 \text{e}^{-2\Gamma t}}-\frac{y^2}{b_0 \text{e}^{2\Gamma t}}  \end{array}
\end{equation}
of the RMHD equations (\ref{RMHDeqn1}) and (\ref{RMHDeqn2}) derived in \cite{chapman_kendall}.  Here the arbitrary rate parameter $\Gamma$, which is set by boundary conditions, is assumed to be fast compared to any diffusion time scale.  This solution corresponds to a thinning and elongating magnetic neutral line at $x=0$, as would be found at the center of a localized collapsing current sheet \cite{waelbroeck89, waelbroeck93, Loureiro08}.
 
This is a particularly relevant solution for the ZS model, as the orderings were constructed with magnetic reconnection studies in mind.  For example, a prototypical model problem would be a localized thinning current sheet whose evolution is eventually disrupted by a reconnecting instability~\citep{UzdenskyLoureiro}.  Typically, in a high temperature plasma, the length scales associated with the reconnecting instability are much smaller than the width of the current sheet itself.  In this circumstance, a localized model of the current sheet such as (\ref{ck_soln}) can capture the salient features of the background which play a role in the physics of the instability, and subsequent nonlinear evolution.   

Using very simple solutions of (\ref{Theta_eqn}), one can generate more exotic versions of the Chapman-Kendall solution.  For example, by choosing $\Theta = z/z_0$, the initial solution (\ref{ck_soln}) transforms to
\begin{equation} \label{transformed_ck}
\begin{array}{cc}\tilde{\phi} = \Gamma \bar{x} \bar{y},& \tilde{\psi} = \bar{x}^2\left(\frac{1}{a_0 \text{e}^{-2\Gamma t}}+\frac{1}{2 z_0} \right)-\bar{y}^2\left(\frac{1}{b_0 \text{e}^{2\Gamma t}}- \frac{1}{2 z_0}\right) \end{array},
\end{equation}
where
\begin{equation}
\begin{array}{cc}\bar{x}=x\,\text{cos}(z/z_0) + y\,\text{sin}(z/z_0),& \bar{y}=y\,\text{cos}(z/z_0) - x\,\text{sin}(z/z_0) \end{array}
\end{equation}
are the helically rotated coordinates.  Physically, this transformation corresponds to a linear helical twisting of the original current sheet with a uniform current (amplitude proportional to the helical pitch) added.  Note that if $b_0 < 2 z_0$, then the flux surfaces will initially be hyperbolic, but at a later time $t_{\text{c}}$ when $b_0 \,\mathrm{exp}(\Gamma t_{\text{c}})=2z_0$, they topologically transform to elliptic surfaces.

This three-dimensional magnetic structure is a simple, analytically-tractable model configuration of an evolving three dimensional magnetic structure that will eventually become unstable to reconnection-driven instabilities.  The formation of helical three-dimensional magnetic fields -- and the potential subsequent magnetic reconnection thought to occur in such structures -- is highly relevant for solar flares~\citep{janvier}.
%\begin{figure}
%  \centerline{\includegraphics[height=7cm,width=13cm]{CKtransformed.eps}}% Images in 100% size
%  \caption{Trapped-mode wavenumbers, $kd$, plotted against $a/d$ for
%    three ellipses:\protect\\
%%    ---$\!$---,
%    $b/a=1$; $\cdots$\,$\cdots$, $b/a=1.5$.}
%\label{fig:ka}
%\end{figure}
\begin{figure}
\centering
\begin{minipage}{.5\textwidth}
  \centering
  \includegraphics[width=.6\linewidth]{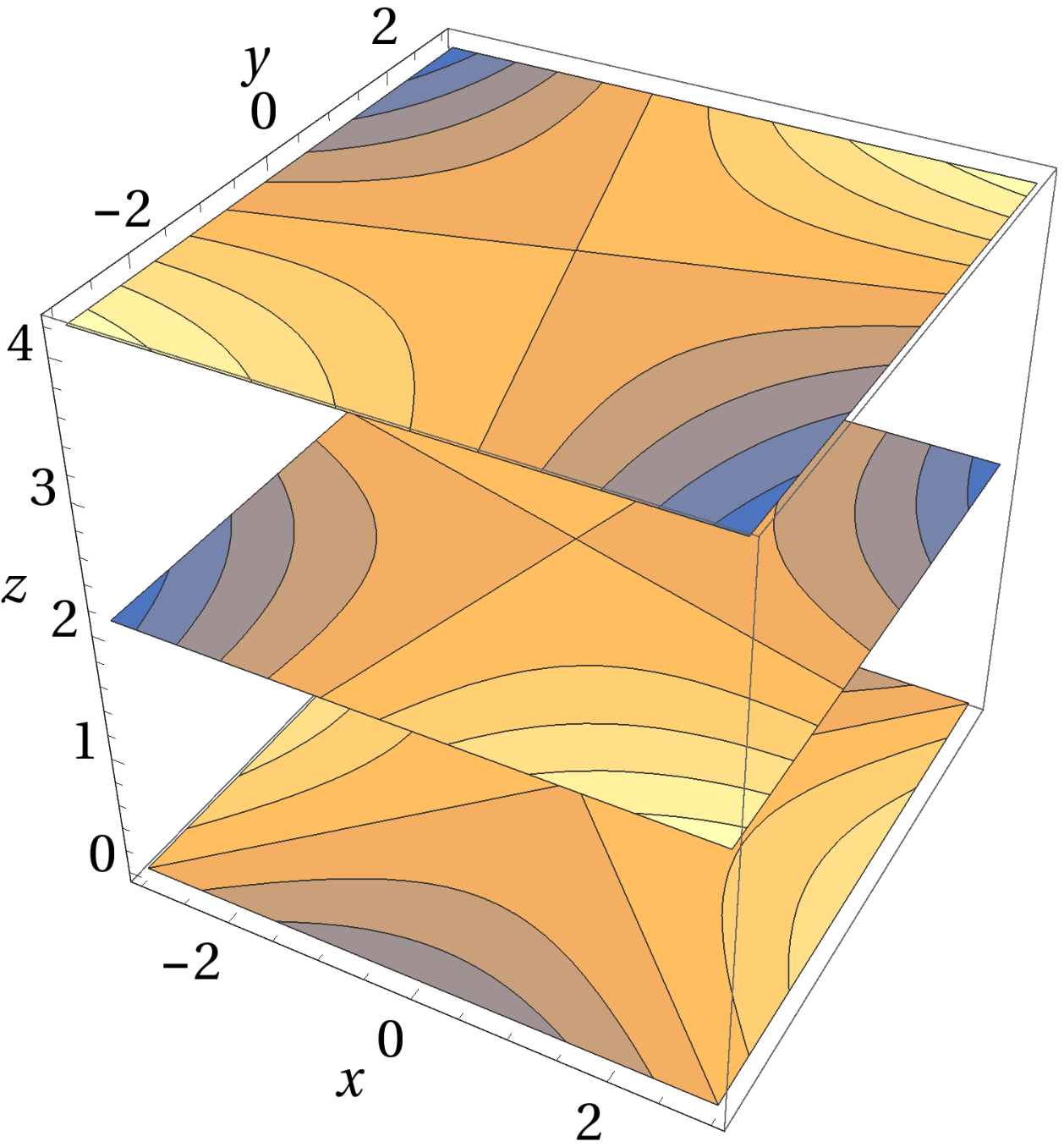}
  %\captionof{figure}{A figure}
  \label{fig:test1}
\end{minipage}%
\begin{minipage}{.5\textwidth}
  \centering
  \includegraphics[width=.6\linewidth]{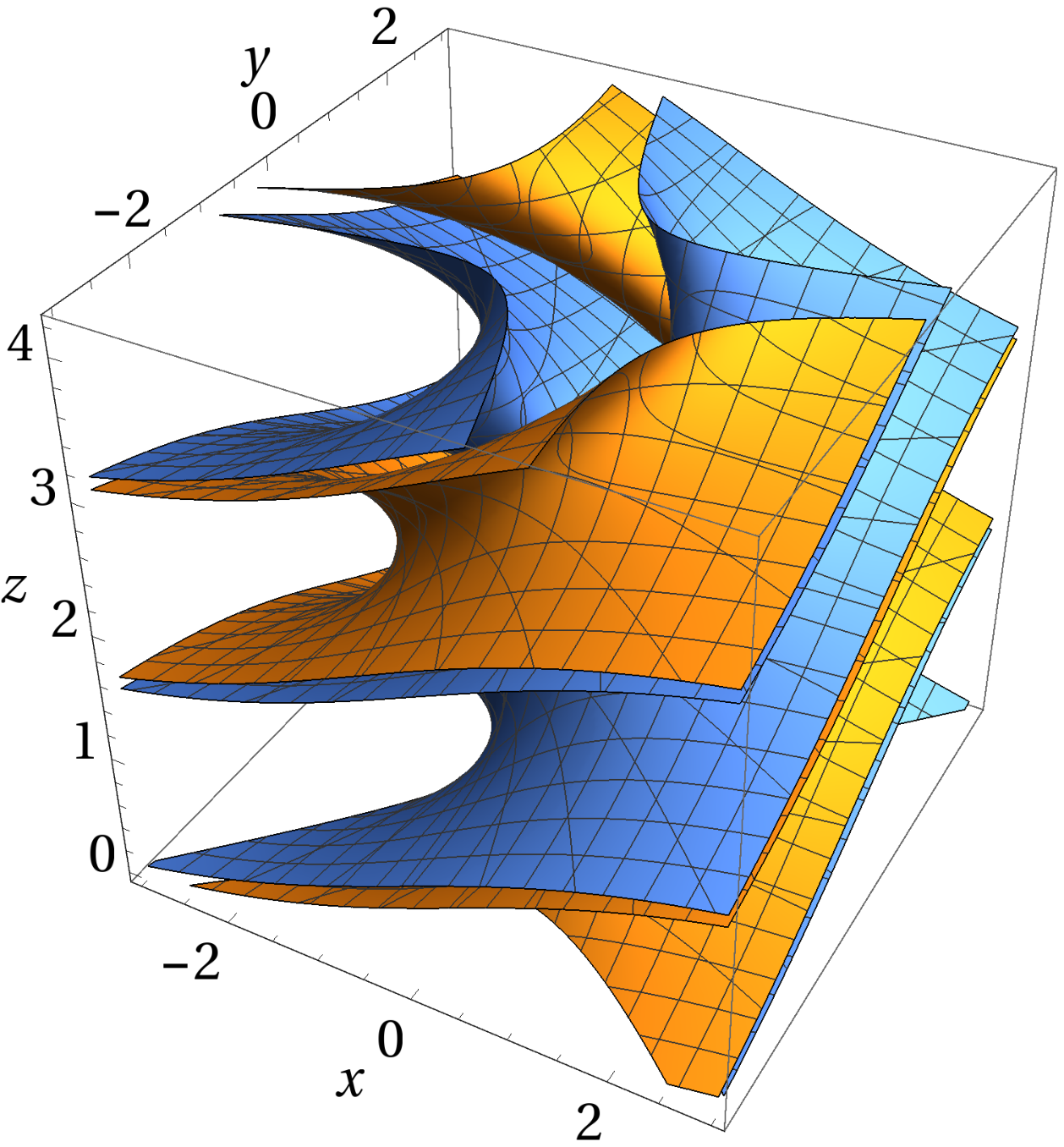}
  %\caption{Flux surface near magnetic null for (\ref{transformed_ck}) with $t=0$, $a= 1$, $b=0.5$ and $z_0 = 1$.}
  \label{fig:test2}
\end{minipage}
\caption{Flux surfaces near magnetic null for (\ref{transformed_ck}) with $t=0$, $a= 1$, $b=0.5$, and $z_0 = 1$.}
\end{figure}
\begin{figure}
  \centerline{\includegraphics[width=0.3\linewidth]{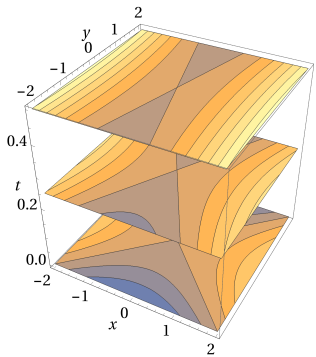}}% Images in 100% size
  \caption{Flux surfaces near magnetic null for (\ref{transformed_ck}) with $z=0$, $a= 1$, $b=0.5$,  $\Gamma = 1$ and $z_0 = 1$.}
%\label{fig:ka}
\end{figure}

As a second example, by choosing $\Theta = (z-v_At)/z_0$, we find a solution which, at time $t=0$, is in the same helical configuration as (\ref{transformed_ck}), but now moves along the guide field at the Alfv\`en speed.  

\section{Symmetry Analysis}\label{analysis}
\subsection{Infinitesimal Generators}
The maps given in section \ref{symmetries_section} can be viewed geometrically as a continuous family of transformations of the graph of the solution, which is a set of points in the (10 dimensional) space labeled by $(v,x,y,z,t; \phi,\psi,\delta n, \delta T, g)$.  These subgroups of transformations can be characterized by their \emph{infinitesimal generators} $\boldsymbol{\Xi}$.

For example, consider the action of an infinitesimal gauge transformation $\delta H(z,t)$ on the graph of a solution $(\phi,\psi,\delta n, \delta T, g)$:
\begin{equation}
(\tilde{v},\tilde{\mathbf{x}}_{\perp},\tilde{z},\tilde{t},\tilde{\phi},\tilde{\psi},\tilde{\delta n}, \tilde{\delta T}, \tilde{g})=[1 - \delta H_t\partial_\phi + \delta H_z\partial_\psi+\mathcal{O}(\delta^2)](v,\mathbf{x},z,t; \phi,\psi,\delta n, \delta T, g).
\end{equation}     
Here, the infinitesimal generator is seen to be
\begin{equation}
\boldsymbol{\Xi}_H=-  H_t\partial_\phi +  H_z\partial_\psi.
\end{equation}
Similarly, the infinitesimal generators for (\ref{perp_translations}) and (\ref{alfven_rotations}) are
\begin{equation}
\boldsymbol{\Xi}_{\boldsymbol{\xi}}=-  \boldsymbol{\xi}\cdot\partial_{\mathbf{x}_{\perp}} + \partial_z \hat{\mathbf{z}}\cdot\boldsymbol{\xi}\times\mathbf{x}_{\perp}\partial_\psi -\partial_t \hat{\mathbf{z}}\cdot\boldsymbol{\xi}\times\mathbf{x}_{\perp}\partial_\phi,
\end{equation}
and
\begin{equation}
\begin{split}
\boldsymbol{\Xi}_\Theta=&\Theta \partial_{\theta} + \partial_z \left(\frac{r^2}{2}\Theta\right)\partial_\psi+\left[ -\partial_t \left(\frac{r^2}{2}\Theta\right)+\mathcal{F}\right]\partial_\phi\\
&+\left[-v_A^{-2}\Delta\partial_t\left(\frac{r^2}{2}\Theta\right)\right]\partial_{\delta n}+\mathcal{T}\partial_{\delta T}+G\partial_{g},
\end{split}
\end{equation}
respectively.

There is a one-to-one correspondence between Lie-group transformations and their infinitesimal generators..  For a rigorous but readable introduction to this formalism, see Chapter 1 of \cite{olver}.  

\subsection{Generators acting on differential equations}
As an extension to a standard graph, one can take a solution of the ZS model and produce a graph in the higher dimensional space consisting of the independent and dependent variables, as well as all higher derivatives up to second order (the highest order that appears in the model equations).  In this higher dimensional \emph{jet} space, labeled $(v,x,y,z,t; \phi,\psi,\delta n, \delta T, g;\phi_v,\,\dots,\phi_t,\,\dots)\equiv(x^i,u^\alpha,u^\alpha_i,u^\alpha_{i,j})$, the action of a symmetry transformation will also involve the coordinates associated with the derivatives:
\[\label{generic_generator}
\boldsymbol{\Xi}_* = \underbrace{\xi^{i}(x,u)\p_{x^{i}} + U^{\alpha}(x,u)\p_{u^{\alpha}}}_{=\boldsymbol{\Xi}} +\sum_{i,j}U^{\alpha;i,j}(x,u)\p_{u^{\alpha}_{ij}}.
\]
  In this higher dimensional space, the model equations, generically expressed in the form
\begin{equation}
F(x^i,u^\alpha,u^\alpha_i,u^\alpha_{i,j})=0\text{ for all } x^i,
\end{equation}
are formally algebraic equations.  

A transformation generated by (\ref{generic_generator}) is a symmetry of the model if the model equations themselves are invariant under the transformation while transforming a solution.  That is
\begin{equation} \label{invar_crit}
\boldsymbol{\Xi}_{*} F = 0\text{ whenever } F=0 \Leftrightarrow \boldsymbol{\Xi}\text{ is a symmetry of }F.
\end{equation}
This \emph{invariance criterion} is a fundamental theorem in symmetry analysis of differential equations.

The second key result  is the \emph{prolongation formula}, which expresses the functions $U^{\alpha;i,j}$ in terms of the $U^{\alpha}$ and $\xi^{i}$.  In other words, given an infinitesimal generator $\boldsymbol{\Xi}$ in the space $(x^i,u^\alpha)$, the prolongation formula provides an explicit expression of the form of the generator $\boldsymbol{\Xi}_*$ in the jet space $(x^i,u^\alpha, u^\alpha_i,u^\alpha_{ij})$.

Using the prolongation formula, the invariance criterion (\ref{invar_crit}) becomes a working procedure to obtain symmetries.  Starting from the general symmetry generator
\begin{equation} \label{generic_Xi}
\boldsymbol{\Xi}=\xi^{i}(x,u)\p_{x^{i}} + U^{\alpha}(x,u)\p_{u^{\alpha}},
\end{equation}
with unknown coefficents $\xi^{i}$ and $U^{\alpha}$, one uses the prolongation formula to compute the form of this symmetry generator (denoted here as $\boldsymbol{\Xi}_{*}$) in jet space.  Once this is computed, the invariance criterion imposes conditions on the coefficients of $\boldsymbol{\Xi}_{*}$ which must be satisfied in order for (\ref{generic_Xi}) to correspond to a symmetry of the system.  By solving these \emph{determining equations}, the most general symmetry of the form (\ref{generic_Xi}) is obtained.

There are two characteristics of this procedure that will be leveraged to extend the procedure to the ZS model, which also has integral relations.  First, in a model consisting of more than one equation, the invariance criterion can be applied to one equation at a time.  The resulting symmetry group will be a subset of the full group; after all, any symmetry of the full model must leave each of its equations unchanged.  Thus we can find the symmetries of the under-determined system (\ref{eqn1})--(\ref{eqn3}) before considering the integral relations (\ref{closure_eqn1}) and (\ref{closure_eqn2}).

The second noteworthy point is that determining equations are usually straightforward to solve, even for highly complicated nonlinear models, provided there are no integral terms.  This justifies the operation of deriving the (generally more complicated) symmetry group of the under-determined model first.  Once the symmetry group of the under-determined model is obtained, this class of transformations is used to simplify application of the integral constraints.

\subsection{Determining equations}
We begin our analysis with a generator of the form
\begin{equation} \label{generator_ansatz}
\boldsymbol{\Xi}=c_1\p_t + c_2\p_z + X\p_{x} + Y\p_{y} +\Phi \p_{\phi} + \Psi \p_{\psi} + G\p_{g} + \mathcal{T} \p_{\delta T}+ N\p_{\delta n},
\end{equation}
where $c_i$ are constants, and the remaining unknown functions depend only on the independent variables.  This is not the most generic form.  Our motivation for choosing this simpler but less general form is based on exploratory computational analysis, using software provided by \cite{cantwell}.  This exploration suggests that all of the symmetries are of the form (\ref{generator_ansatz}).  

\subsubsection{Local determining equations}
Following the procedure sketched in the preceding subsection, we obtain from (\ref{eqn1}) and (\ref{eqn3}) the following determining equations:
\begin{eqnarray}
X_v &=& Y_v = \Phi_v = \Psi_v = T_v = N_v = 0,\label{dt0}\\
X_{x}+ Y_{y}  &=& 0, \label{dt1}\\
X_{x} &=& Y_{y} = 0, \label{dt2}\\
\Psi_{x} -Y_{z} &=& \Psi_{y} - X_{z} = 0, \label{dt3}\\
\Phi_{y} + X_{t}  &=&\Phi_{x} - Y_{t}  = 0, \label{dt4}\\
%\nabla_{\perp}^{2} X &=&\nabla_{\perp}^{2} Y = 0 \label{dt5} \\
N_x &=& N_{y} = 0, \label{dt6}\\
\mathcal{T}_{x} &=& \mathcal{T}_{y} = 0, \label{dt7}\\
G_{x}&=& G_{y}  = 0,  \label{dt8}\\
N_{t} &=& \Delta \Psi_{z}, \label{dt9}\\
\Psi_{t} - \Delta \Psi_{t} &=& \Delta \Psi - \Phi_{z} - N_{z} - \mathcal{T}_{z}, \label{dt10}\\
G_{t} + v(G_{z} +\mathcal{T}) &=& G_{vv} - vG_{v} \nonumber \\ &+& (1-v^{2})(N_{t}+\mathcal{T}).\label{dt11}
\end{eqnarray}
These linear differential equations are sometimes called the ``local'' determining equations; symmetries of the non-local, integral relations (\ref{closure_eqn1}) -- (\ref{closure_eqn2}) remain to be considered.

\subsubsection{Integral determining equations}
For integrodifferential equations, the notion of the jet space (itself a generalization of the graph space) can be extended to include the moments of independent variables which appear in the model.  We denote the variables in this space generically as $(x^i,u^\alpha,u^\alpha_i,u^\alpha_{i,j},m^{\mu}\equiv\int K^{\mu}_\alpha(x,x')u^\alpha(x'))$.  For the ZS model, the two moment variables are $\delta n$ and $\delta T$.  What is needed for the invariance criterion (\ref{invar_crit}) is the expression for $\boldsymbol{\Xi}m^{\mu}$ in terms of the generator coefficients $U^{\alpha}$ and $X^{i}$. 

For this purpose, there is a very useful fact: one can re-express $\boldsymbol{\Xi}$ in \emph{canonical form}, where it acts only on the dependent variables:
\begin{equation}\label{canon_xi}
\boldsymbol{\Xi}=\sum_{\alpha}Q^\alpha \partial_{u^{\alpha}},
\end{equation}
with
\begin{equation}
Q^{\alpha} = U^\alpha-\sum_i X^{i}u^{\alpha}_{i}.
\end{equation}
It turns out \cite[see e.g.][]{kovalev} that the action of the canonical generator (\ref{canon_xi}) on an integral -- a functional of the $u^\alpha$ -- is obtained by replacing ordinary derivatives by functional derivatives in the canonical expression
\begin{equation}\label{action_on_moment}
\boldsymbol{\Xi}m^{\mu} = \sum_{\alpha}Q^\alpha \frac{\delta m^{\mu}}{\delta u^{\alpha}}.
\end{equation}

Using (\ref{action_on_moment}), the invariance criteria $\boldsymbol{\Xi}$(\ref{closure_eqn1})$=0$ and $\boldsymbol{\Xi}$(\ref{closure_eqn2})$=0$ give the \emph{integral determining equations}
\begin{equation}\label{ide_1}
\mathcal{T} = \frac{1}{\sqrt{2\pi}}\int\mathrm{d}v' v'^2 \mathrm{e}^{-{v'}^2/2}G(v'),\\
\end{equation}
and
\begin{equation}\label{ide_2}
N = \hat{\mathcal{G}}\Phi,
\end{equation}
respectively.  Similarly, the integral constraints (\ref{norm_int_constraints}) lead to the determining equations
\begin{equation} \label{ide_3}
\left(\begin{array}{c}0\\0\end{array}\right) = \frac{1}{\sqrt{2\pi}}\int\mathrm{d}v' \left(\begin{array}{c}0\\v'\end{array}\right) \mathrm{e}^{-{v'}^2/2}G(v').
\end{equation}

Thus our full system of equations for the generator coefficients consist of the local equations (\ref{dt1})--(\ref{dt11}) together with the integral relations (\ref{ide_1})--(\ref{ide_3}).  Note in particular that this system of equations also involves a gyroaveraging operator, as well as other integral relations which usually lead to analytical intractability.  However in this case, the local determining equations form an autonomous subsystem; we are able to obtain their general solution before even deriving the remaining (integral) determining equations.  

\subsection{Solution of determining equations} \label{soln_DEs}
The general solution of the local determining equations (\ref{dt0})--(\ref{dt11}) is
\begin{eqnarray}
X &=& \Theta y + \hat{\mathbf{x}}\cdot\boldsymbol{\xi}, \label{finfirst}\\
Y &=& - \Theta x + \hat{\mathbf{y}}\cdot\boldsymbol{\xi},\\
\Psi &=& \partial_z\left(\Theta\frac{x^2 + y^2}{2} +\hat{\mathbf{z}}\cdot\boldsymbol{\xi}\times\mathbf{x}_{\perp} + H\right),\\
\Phi &=& -\partial_t \left(\Theta\frac{x^2 + y^2}{2} +\hat{\mathbf{z}}\cdot\boldsymbol{\xi}\times\mathbf{x}_{\perp} + H\right) + \left[ 2(\Theta_{t}+\Theta) +\mathcal{T}+N\right],\label{solnPhi}\\
N &=& -2v_A^{-2}\Theta_t\\
\mathcal{T} &=& \mathcal{T}(z,t),\\
 G_t +v G_z &=&-G_{vv} +v G_v -(1-v^2)(N_t + \mathcal{T}) - vT_z,\label{Gde}
\end{eqnarray}
Here $\hat{\mathbf{x}}\cdot\boldsymbol{\xi}(z,t)$, $\hat{\mathbf{y}}\cdot\boldsymbol{\xi}(z,t)$, and $H(z,t)$ are arbitrary functions, while $\Theta(z,t)$ is an arbitrary solution to the wave equation (\ref{Theta_eqn}).

In obtaining this result, we have used
\begin{align}
\hat{\mathcal{G}}\Phi =&  \frac{1}{v_A^2}\Delta \Phi,\\
                =&-\frac{2}{v_A^2}\Theta_t.
\end{align}
In other words, the leading FLR approximation to $\hat{\mathcal{G}}$, shown in (\ref{FLRgyro}), here becomes exact, since the $\Phi$ given in (\ref{solnPhi}) is quadratic in the perpendicular coordinates.

The function $G$ is determined implicitly by (\ref{Gde}) and constrained to satisfy
\begin{eqnarray}
0 &=& \int \text{d}v' G(v')\text{e}^{-{v'}^2/2},\\
0 &=& \int \text{d}v' v'G(v')\text{e}^{-{v'}^2/2}. \label{finlast}
\end{eqnarray}

The symmetries discussed in Section \ref{symmetries_section} follow from Eqs. (\ref{finfirst})--(\ref{finlast}).

\section{Conclusion}\label{con}
We have found that the fluid-gyrokinetic ZS model has a rich symmetry group, with the full set of symmetries spanned by five arbitrary functions $H(z,t)$, $\hat{\mathbf{x}}\cdot\boldsymbol{\xi}(z,t)$, $\hat{\mathbf{y}}\cdot\boldsymbol{\xi}(z,t)$, and $\Theta^{\pm}(z\pm v_A t)$.  These symmetries are discussed in Section \ref{symmetries_section}, and summarized by Eqs. (\ref{generator_ansatz}), with (\ref{finfirst})--(\ref{finlast}). 

To our knowledge, this is the first time symmetry analysis has been applied to a model with a gyroaveraging operator.  Gyroaveraging, viewed as a constitutive relation linking $\phi$ to $\delta n$, turns out to not pose a serious obstacle in our analysis, largely because the $x_{\perp}$ dependence of the infinitesimal generator for $\phi$ allowed the exact gyroaverage to be expressed in closed form.  Similarly, the generator $G$ for displacement symmetry of the distribution function is found to satisfy exactly a \emph{linear} drift-kinetic equation.  The success of symmetry analysis for the ZS model suggests that a similar study for the full gyrokinetic equations might also be possible.  

Our analysis assumes, based on computational exploration, a special form for the symmetries, so it is possible that additional symmetries remain undiscovered.  In fact, even if we did begin our analysis with the most general possible transformation, the integral terms in ZS place it beyond the scope of the theorems that would prove completeness.

Because RMHD is a limit of the ZS model, our analysis also provides a large family of symmetries of RMHD.  RMHD is a simpler and better studied model, so there are more exact solutions available to transform by our methods.  In particular, the results obtained here can be used to generate new exact solutions to RMHD by transforming the Elsasser solutions \citep{elsasser}, which play an important role in MHD theories of turbulence \cite[for a review, see, e.g.,][]{biskamp}.

For illustrative purposes, the modified Chapman-Kendall solution obtained in Section \ref{trans_ck} employed a very simple particular symmetry transformation.  More generally, using the full set of transformations obtained here, the original two parameter Chapman-Kendall solution becomes a large family of solutions, spanned by the arbitrary functions $H$, $\boldsymbol{\xi}$ and $\Theta$.  

In the context of the full kinetic model, one can, for example, leverage simulation results that start from a Chapman-Kendall-like two-dimensional current configuration to infer the behavior of a whole family of (generally three dimensional) initial current profiles, such as the helical collapsing current sheet given in (\ref{transformed_ck}). 

Noether's theorem applies to all of the symmetry transformations obtained here.  If one is able to construct an action for this model~\citep[see][for manifestly action-preserving derivations of reduced models]{Ioannis, Morrison, burby}, and if the action is invariant under any of these transformations, then one can use Noether's theorem to derive conserved quantities which, like the symmetries themselves, may not be obvious from physical considerations.  Symmetry analysis can thus enhance the value of a reduced model by uncovering quantities which, while perhaps not exactly conserved in the full Maxwell-Boltzmann description, are approximately constant in particular regimes of interest.  For the ZS model, this context would be nonlinear fluctuations in a high temperature strongly-magnetized plasma.

\section*{Acknowledgements}
R.L.W. was supported by This research was supported by the U.S. Department of Energy Fusion Energy Sciences Postdoctoral Research Program administered by the Oak Ridge Institute for Science and Education (ORISE) for the DOE. ORISE is managed by Oak Ridge Associated Universities (ORAU) under DOE contract number DE-SC0014664. All opinions expressed in this paper are the author's and do not necessarily reflect the policies and views of DOE, ORAU, or ORISE.  The work of R.D.H. was funded by the U.S. Department of Energy under Contract No. DE-FG02-04ER-54742 and by The University of Texas at Austin.  N.F.L. was partially funded by US Department of Energy Grant No. DE-FG02-91ER54109.  We also acknowledge the use of Brian Cantwell's Lie-group software, implemented on \mathematica.
\bibliographystyle{jpp}
\bibliography{jppSymmetryZS_final}

\begin{thebibliography}{51}
\expandafter\ifx\csname natexlab\endcsname\relax\def\natexlab#1{#1}\fi
\def\au#1{#1} \def\ed#1{#1} \def\yr#1{#1}\def\at#1{#1}\def\jt#1{\textit{#1}}
  \def\bt#1{#1}\def\bvol#1{\textbf{#1}} \def\vol#1{#1} \def\pg#1{#1}
  \def\publ#1{#1}\def\arxiv#1{#1}\def\org#1{#1}\def\st#1{\textit{#1}}

\bibitem[Beer \& Hammett(1996)]{Beer96}
{\sc \au{Beer, M.} \& \au{Hammett, G.~W.}} \yr{1996}  \at{{Toroidal gyrofluid
  equations for simulations of tokamak turbulence}}.  \jt{Phys. Plasmas}
  \bvol{3},  \pg{4046}.

\bibitem[Bian \& Kontar(2010)]{bian}
{\sc \au{Bian, N.~H.} \& \au{Kontar, E.~P.}} \yr{2010}  \at{{A gyrofluid
  description of Alfv\'enic turbulence and its parallel electric field}}.
  \jt{Phys. Plasmas}  \bvol{17},  \pg{062308}.

\bibitem[Biskamp(2003)]{biskamp}
{\sc \au{Biskamp, D.}} \yr{2003} {\em Magnetohydrodynamic Turbulence\/}.
  \publ{Cambridge: Cambridge University Press}.

\bibitem[Brizard(1992)]{Brizard92}
{\sc \au{Brizard, A.}} \yr{1992}  \at{{Nonlinear gyrofluid description of
  turbulent magnetized plasmas}}.  \jt{Phys. Fluids B}  \bvol{4},  \pg{1213}.

\bibitem[Burby(2017)]{burby}
{\sc \au{Burby, J.~W.}} \yr{2017}  \at{Magnetohydrodynamic motion of a
  two-fluid plasma}.  \jt{Physics of Plasmas}  \bvol{24}~(8),  \pg{082104}.

\bibitem[Cantwell(2002)]{cantwell}
{\sc \au{Cantwell, B.~J.}} \yr{2002} {\em Introduction to symmetry analysis\/}.
   \publ{Cambridge: Cambridge University Press}.

\bibitem[Catto(1978)]{CattoI}
{\sc \au{Catto, P.~J.}} \yr{1978}  \at{{Linearized gyro-kinetics}}.  \jt{Plasma
  Phys.}  \bvol{20},  \pg{719}.

\bibitem[Catto {\em et~al.\/}(1981)Catto, Tang \& Baldwin]{CattoII}
{\sc \au{Catto, P.~J.}, \au{Tang, W.~M.} \& \au{Baldwin, D.~E.}} \yr{1981}
  \at{{Generalized gyrokinetics}}.  \jt{Plasma Phys.}  \bvol{23},  \pg{639}.

\bibitem[Chapman \& Kendall(1963)]{chapman_kendall}
{\sc \au{Chapman, S.} \& \au{Kendall, P.~C.}} \yr{1963}  \at{Liquid instability
  and energy transformation near a magnetic neutral line: a soluble nonlinear
  hydromagnetic problem}.  \jt{Proc. Roy. Soc. A}  \bvol{271},  \pg{435--448}.

\bibitem[Charidakos {\em et~al.\/}(2014)Charidakos, Lingam, Morrison, White \&
  Wurm]{Ioannis}
{\sc \au{Charidakos, I.~K.}, \au{Lingam, M.}, \au{Morrison, P.~J.}, \au{White,
  R.~L.} \& \au{Wurm, A.}} \yr{2014}  \at{Action principles for extended
  magnetohydrodynamic models}.  \jt{Phys. Plasmas}  \bvol{21}~(9),
  \pg{092118}.

\bibitem[Connor \& Taylor(1977)]{ConnnorTaylor1977}
{\sc \au{Connor, J.~W.} \& \au{Taylor, J.~B.}} \yr{1977}  \at{Scaling laws for
  plasma confinement}.  \jt{Nucl. Fusion}  \bvol{17}~(5),  \pg{1047}.

\bibitem[Dimits {\em et~al.\/}(2000)Dimits, Bateman, Beer, Cohen, Dorland,
  Hammett, Kim, Kinsey, Kotschenreuther, Kritz, Lao, Mandrekas, Nevins, Parker,
  Redd, Shumaker, Sydora \& Weiland]{Dimits}
{\sc \au{Dimits, A.~M.}, \au{Bateman, G.}, \au{Beer, M.~A.}, \au{Cohen, B.~I.},
  \au{Dorland, W.}, \au{Hammett, G.~W.}, \au{Kim, C.}, \au{Kinsey, J.~E.},
  \au{Kotschenreuther, M.}, \au{Kritz, A.~H.}, \au{Lao, L.~L.}, \au{Mandrekas,
  J.}, \au{Nevins, W.~M.}, \au{Parker, S.~E.}, \au{Redd, A.~J.}, \au{Shumaker,
  D.~E.}, \au{Sydora, R.} \& \au{Weiland, J.}} \yr{2000}  \at{Comparisons and
  physics basis of tokamak transport models and turbulence simulations}.
  \jt{Phys. Plasmas}  \bvol{7}~(3),  \pg{969--983}.

\bibitem[Dorland \& Hammett(1993)]{DorlandHammett93}
{\sc \au{Dorland, W.} \& \au{Hammett, G.~W.}} \yr{1993}  \at{{Gyrofluid
  turbulence models with kinetic effects}}.  \jt{Phys. Fluids B}  \bvol{5},
  \pg{812}.

\bibitem[Dubin {\em et~al.\/}(1983)Dubin, Krommes, Oberman \& Lee]{Dubin}
{\sc \au{Dubin, D.~H.}, \au{Krommes, J.~A.}, \au{Oberman, C.} \& \au{Lee,
  W.~W.}} \yr{1983}  \at{{Nonlinear gyrokinetic equations}}.  \jt{Phys. Fluids}
   \bvol{26},  \pg{3524}.

\bibitem[Elsasser(1950)]{elsasser}
{\sc \au{Elsasser, W.~M.}} \yr{1950}  \at{The hydromagnetic equations}.
  \jt{Phys. Rev.}  \bvol{79},  \pg{183}.

\bibitem[Frieman \& Chen(1982)]{FriemanChen}
{\sc \au{Frieman, E.~A.} \& \au{Chen, L.}} \yr{1982}  \at{{Nonlinear
  gyrokinetic equations for low-frequency electromagnetic waves in general
  plasma equilibria}}.  \jt{Phys. Fluids}  \bvol{25},  \pg{502}.

\bibitem[{Hahm} {\em et~al.\/}(1988){Hahm}, {Lee} \& {Brizard}]{HahmLee}
{\sc \au{{Hahm}, T.~S.}, \au{{Lee}, W.~W.} \& \au{{Brizard}, A.}} \yr{1988}
  \at{{Nonlinear gyrokinetic theory for finite-beta plasmas}}.  \jt{Phys.
  Fluids}  \bvol{31},  \pg{1940--1948}.

\bibitem[Hammett {\em et~al.\/}(1993)Hammett, Beer, Dorland, Cowley \&
  Smith]{HammettEtal93}
{\sc \au{Hammett, G.~W.}, \au{Beer, M.}, \au{Dorland, W.}, \au{Cowley, S.~C.}
  \& \au{Smith, S.~A.}} \yr{1993}  \at{{Developments in the gyrofluid approach
  to tokamak turbulence simulations}}.  \jt{Plasma Phys. Control. Fusion}
  \bvol{35},  \pg{973}.

\bibitem[Hammett {\em et~al.\/}(1992)Hammett, Dorland \& Perkins]{Hammett92}
{\sc \au{Hammett, G.~W.}, \au{Dorland, W.} \& \au{Perkins, F.~W.}} \yr{1992}
  \at{{Fluid models of phase mixing, Landau damping, and nonlinear gyrokinetic
  dynamics}}.  \jt{Phys. Fluids B}  \bvol{4},  \pg{2052}.

\bibitem[Hammett \& Perkins(1990)]{HammettPerkins}
{\sc \au{Hammett, G.~W.} \& \au{Perkins, F.~W.}} \yr{1990}  \at{{Fluid models
  for Landau damping with application to the ion-temperature-gradient
  instability}}.  \jt{Phys. Rev. Lett.}  \bvol{64},  \pg{3019}.

\bibitem[Hatch {\em et~al.\/}(2014)Hatch, Jenko, Bratanov \& {Banon
  Navarro}]{hatch2014}
{\sc \au{Hatch, D.~R.}, \au{Jenko, F.}, \au{Bratanov, V.} \& \au{{Banon
  Navarro}, A.}} \yr{2014}  \at{Phase space scales of free energy dissipation
  in gradient-driven gyrokinetic turbulence}.  \jt{J. Plasma Phys.}  \bvol{80},
   \pg{531}.

\bibitem[{Janvier} {\em et~al.\/}(2013){Janvier}, {Aulanier}, {Pariat} \&
  {D{\'e}moulin}]{janvier}
{\sc \au{{Janvier}, M.}, \au{{Aulanier}, G.}, \au{{Pariat}, E.} \&
  \au{{D{\'e}moulin}, P.}} \yr{2013}  \at{{The standard flare model in three
  dimensions. III. Slip-running reconnection properties}}.  \jt{A\&A}
  \bvol{555},  \pg{A77}.

\bibitem[Kadomtsev \& Pogutse(1974)]{kadomtsev}
{\sc \au{Kadomtsev, B.~B.} \& \au{Pogutse, O.~P.}} \yr{1974}  \at{{Nonlinear
  helical perturbations of a plasma in the tokamak}}.  \jt{Soviet Phys. JETP}
  \bvol{38},  \pg{283}.

\bibitem[Kanekar {\em et~al.\/}(2015)Kanekar, Schekochihin, Dorland \&
  Loureiro]{kanekar_schekochihin_dorland_loureiro_2015}
{\sc \au{Kanekar, A.}, \au{Schekochihin, A.~A.}, \au{Dorland, W.} \&
  \au{Loureiro, N.~F.}} \yr{2015}  \at{Fluctuation-dissipation relations for a
  plasma-kinetic langevin equation}.  \jt{J. Plasma Phys.}  \bvol{81}~(1).

\bibitem[Kovalev {\em et~al.\/}(1996)Kovalev, Krivenko \& Pustovalev]{kovalev}
{\sc \au{Kovalev, V.~F.}, \au{Krivenko, S.~V.} \& \au{Pustovalev, V.}}
  \yr{1996}  \at{{Symmetry group of Maxwell-Vlasov equations in plasma
  theory}}.  \jt{‎J. Nonlinear Math. Phys.}  \bvol{3}~(1-2),  \pg{175--180}.

\bibitem[Kruskal \& Oberman(1958)]{kruskaloberman}
{\sc \au{Kruskal, M.~D.} \& \au{Oberman, C.~R.}} \yr{1958}  \at{{On the
  Stability of Plasma in Static Equilibrium}}.  \jt{Phys. Fluids}
  \bvol{1}~(4),  \pg{275}.

\bibitem[Lee(1983)]{LeeI}
{\sc \au{Lee, W.~W.}} \yr{1983}  \at{{Gyrokinetic approach in particle
  simulation}}.  \jt{Phys. Fluids}  \bvol{26},  \pg{555}.

\bibitem[Lee(1987)]{LeeII}
{\sc \au{Lee, W.~W.}} \yr{1987}  \at{{Gyrokinetic particle simulation model}}.
  \jt{J. Comput. Phys.}  \bvol{72},  \pg{243}.

\bibitem[Loureiro {\em et~al.\/}(2005)Loureiro, Cowley, Dorland, Haines \&
  Schekochihin]{Loureiro08}
{\sc \au{Loureiro, N.~F.}, \au{Cowley, S.~C.}, \au{Dorland, W.~D.}, \au{Haines,
  M.~G.} \& \au{Schekochihin, A.~A.}} \yr{2005}  \at{X-point collapse and
  saturation in the nonlinear tearing mode reconnection}.  \jt{Phys. Rev.
  Lett.}  \bvol{95},  \pg{235003}.

\bibitem[Loureiro {\em et~al.\/}(2016)Loureiro, Dorland, Fazendeiro, Kanekar,
  Mallet, Vilelas \& Zocco]{viriato}
{\sc \au{Loureiro, N.~F.}, \au{Dorland, W.}, \au{Fazendeiro, L.}, \au{Kanekar,
  A.}, \au{Mallet, A.}, \au{Vilelas, M.~S.} \& \au{Zocco, A.}} \yr{2016}
  \at{{Viriato: a Fourier-Hermite spectral code for strongly magnetized
  fluid-kinetic plasma dynamics}}.  \jt{Comput. Phys. Commun.}  \bvol{206},
  \pg{45}.

\bibitem[Morrison {\em et~al.\/}(2014)Morrison, Lingam \& Acevedo]{Morrison}
{\sc \au{Morrison, P.~J.}, \au{Lingam, M.} \& \au{Acevedo, R.}} \yr{2014}
  \at{Hamiltonian and action formalisms for two-dimensional gyroviscous
  magnetohydrodynamics}.  \jt{Phys. Plasmas}  \bvol{21}~(8),  \pg{082102}.

\bibitem[Olver(1993)]{olver}
{\sc \au{Olver, P.~J.}} \yr{1993} {\em Applications of Lie Groups to
  Differential Equations\/}.  \publ{Springer-Verlag}.

\bibitem[Ramos(2010)]{RamosI}
{\sc \au{Ramos, J.~J.}} \yr{2010}  \at{{Fluid and drift-kinetic description of
  a magnetized plasma with low collisionality and slow dynamics orderings. I.
  Electron theory}}.  \jt{Phys. Plasmas}  \bvol{17},  \pg{082502}.

\bibitem[Ramos(2011)]{RamosII}
{\sc \au{Ramos, J.~J.}} \yr{2011}  \at{{Fluid and drift-kinetic description of
  a magnetized plasma with low collisionality and slow dynamics orderings. II.
  Ion theory}}.  \jt{Phys. Plasmas}  \bvol{18},  \pg{102506}.

\bibitem[Rayleigh(1891)]{ray}
{\sc \au{Rayleigh, L.}} \yr{1891}  \at{{Dynamical problems in illustration of
  the theory of gases}}.  \jt{Phil. Mag.}  \bvol{32},  \pg{424}.

\bibitem[Roberts({1985})]{Roberts85}
{\sc \au{Roberts, D.}} \yr{{1985}}  \at{{The general Lie group and similarity
  solutions for the one-dimensional Vlasov-Maxwell equations}}.  \jt{{J.~Plasma
  Phys.}}  \bvol{{33}}~({2}),  \pg{{219}}.

\bibitem[Rosenbluth \& Rostoker(1959)]{rosenbluth-rostoker}
{\sc \au{Rosenbluth, M.~N.} \& \au{Rostoker, N.}} \yr{1959}  \at{{Theoretical
  Structure of Plasma Equations}}.  \jt{Phys. Fluids}  \bvol{2}~(1),  \pg{23}.

\bibitem[Rutherford \& Frieman(1968)]{RuthFr}
{\sc \au{Rutherford, P.~H.} \& \au{Frieman, E.~A.}} \yr{1968}  \at{{Drift
  instabilities in general magnetic field configurations}}.  \jt{Phys. Fluids}
  \bvol{11},  \pg{569}.

\bibitem[Schekochihin {\em et~al.\/}(2016)Schekochihin, Parker, Highcock,
  Dellar, Dorland \& Hammett]{scheko2016}
{\sc \au{Schekochihin, A.~A.}, \au{Parker, J.~T.}, \au{Highcock, E.~G.},
  \au{Dellar, P.~J.}, \au{Dorland, W.} \& \au{Hammett, G.~W.}} \yr{2016}
  \at{Phase mixing versus nonlinear advection in drift-kinetic plasma
  turbulence}.  \jt{J. Plasma Phys.}  \bvol{82},  \pg{905820212}.

\bibitem[Snyder \& Hammett(2001)]{Snyder2001}
{\sc \au{Snyder, P.~B.} \& \au{Hammett, G.~W.}} \yr{2001}  \at{{Electromagnetic
  effects on plasma microturbulence}}.  \jt{Phys. Plasmas}  \bvol{8},
  \pg{744}.

\bibitem[Strauss(1976)]{strauss}
{\sc \au{Strauss, H.~R.}} \yr{1976}  \at{Nonlinear, three dimensional
  magnetohydrodynamics of noncircular tokamaks}.  \jt{Phys. Fluids}  \bvol{19},
   \pg{134--140}.

\bibitem[Taylor \& Hastie(1968)]{TaylorHastie}
{\sc \au{Taylor, J.~B.} \& \au{Hastie, R.~J.}} \yr{1968}  \at{{Stability of
  general plasma equilibria-I formal theory}}.  \jt{Phys. Fluids}  \bvol{10},
  \pg{479}.

\bibitem[Uzdensky \& Loureiro(2016)]{UzdenskyLoureiro}
{\sc \au{Uzdensky, D.~A.} \& \au{Loureiro, N.~F.}} \yr{2016}  \at{{Magnetic
  reconnection onset via disruption of a forming current sheet by the tearing
  instability}}.  \jt{Phys. Rev. Lett.}  \bvol{116},  \pg{105003}.

\bibitem[Waelbroeck(1989)]{waelbroeck89}
{\sc \au{Waelbroeck, F.~L.}} \yr{1989}  \at{{Current sheets and the nonlinear
  growth of the m=1 kink-tearing mode}}.  \jt{Phys. Fluids B}  \bvol{1},
  \pg{2372}.

\bibitem[Waelbroeck(1993)]{waelbroeck93}
{\sc \au{Waelbroeck, F.~L.}} \yr{1993}  \at{{Onset of the sawtooth crash}}.
  \jt{Phys. Rev. Lett.}  \bvol{70},  \pg{3259}.

\bibitem[Waelbroeck {\em et~al.\/}(2009)Waelbroeck, Hazeltine \&
  Morrison]{flw09}
{\sc \au{Waelbroeck, F.~L.}, \au{Hazeltine, R.~D.} \& \au{Morrison, P.~J.}}
  \yr{2009}  \at{{A Hamiltonian electromagnetic gyrofluid model}}.  \jt{Phys.
  Plasmas}  \bvol{16},  \pg{032109}.

\bibitem[Watanabe \& Sugama(2004)]{ws2004}
{\sc \au{Watanabe, T.-H.} \& \au{Sugama, H.}} \yr{2004}  \at{Kinetic simulation
  of steady states of ion temperature gradient driven turbulence with weak
  collisionality}.  \jt{Phys. Plasmas}  \bvol{11}~(4),  \pg{1476--1483}.

\bibitem[White \& Hazeltine(2009)]{SymGS}
{\sc \au{White, R.~L.} \& \au{Hazeltine, R.~D.}} \yr{2009}  \at{{Symmetry
  analysis of the Grad-Shafranov equation}}.  \jt{Phys. Plasmas}  \bvol{16},
  \pg{123101}.

\bibitem[White \& Hazeltine(2017)]{rr}
{\sc \au{White, R.~L.} \& \au{Hazeltine, R.~D.}} \yr{2017}  \at{{Analysis of
  the Hermite spectrum in plasma turbulence}}.  \jt{Phys. Plasmas}  \bvol{24},
  \pg{102315}.

\bibitem[Zocco {\em et~al.\/}(2015)Zocco, Loureiro, Dickinson, Numata \&
  Roach]{Zocco2015}
{\sc \au{Zocco, A.}, \au{Loureiro, N.~F.}, \au{Dickinson, D.}, \au{Numata, R.}
  \& \au{Roach, C.~M.}} \yr{2015}  \at{{Kinetic microtearing modes and
  reconnecting modes in strongly magnetised slab plasmas}}.  \jt{Plasma Phys.
  Control. Fusion}  \bvol{57},  \pg{065008}.

\bibitem[Zocco \& Schekochihin(2011)]{zoccoetal}
{\sc \au{Zocco, A.} \& \au{Schekochihin, A.~A.}} \yr{2011}  \at{{Reduced
  fluid-kinetic equations for low-frequency dynamics, magnetic reconnection,
  and electron heating in low-beta plasmas}}.  \jt{Phys. Plasmas}  \bvol{18},
  \pg{102309}.

\end{thebibliography}

\end{document}